# Fragmentation and Dijet Mass Resolution
## Dan Green
### Fermilab

**January, 2004**

## Introduction

The LHC machine is designed to work at a luminosity of $\ell = 10^{34} /(cm^2 \sec)$. The resulting high event rate means that there are ~ 20 minimum bias events in each and every bunch crossing. While the use of tracking information may reduce the pileup from charged particles emanating from uninteresting vertices, it is clear that the remaining "pileup" can potentially degrade the performance of detectors operating at the LHC.

A study has been made of a sequential gauge boson, Z', with a mass of 120 GeV which decays into two jets, or dijets. This process supplies a mass peak, which can be optimized. The width to mass ratio is taken to be the figure of merit when evaluating detector performance at different luminosities. The CMS detector [1] has been fully simulated at a luminosity of 1/5 of design and at design luminosity. The simulated calorimetric information is then used in this particular study.

## QCD and Jet Phase Space

The predictions of QCD [2, 3, 4] in the modified leading log approximation (MLLA) are summarized in Eq. 1.

$$dN/dy = (4/3)[\Gamma(B)/\pi] \int_{-\pi/2}^{\pi/2} d\tau e^{-B\alpha} \{[\cosh\alpha + (1-2p)\sinh\alpha]/[4/3Y(\alpha/\sinh\alpha)]\}^{B/2}$$

$$I_B(\sqrt{(16/3)Y\alpha/\sinh\alpha[\cosh\alpha + (1-2p)\sinh\alpha]})$$

$$y = \log(1/x), \quad x = 2k/M$$
$$Y = \log(2M\sin\theta_c/Q)$$
$$p = 1 - y/Y \qquad\qquad 1)$$
$$\alpha = \alpha_o + i\tau$$
$$\tanh(\alpha_o) = (2p-1)$$
$$B = 101/81$$
$$Q = 0.23\, GeV$$

The multiplicity, N, of partons hadrons within a jet as a function of the momentum fraction of the jet carried by the parton, x, is dN/dy, where $y = \ln(1/x)$. The dijet mass is M. The single parameter is Q, the virtuality where the partons are taken to evolve into hadrons. Local parton-hadron duality (LPD) is assumed where partons become hadrons at mass Q. The CDF collaboration has determined Q to be ~ 0.23 GeV [5, 6].

The y distribution of the mean multiplicity, dN/dy, is predicted to be a function of the single scaling variable $Y = \ln(2M \sin\theta_c / Q)$ where $\theta_c$ is the angle of the "cone" about the initial jet direction. This result is reminiscent of the scaling of electromagnetic cascades in terms of the variable E/Ec, where Ec is the critical energy. The mean multiplicity, N, is the integral of dN/dy over y. As seen in Eq. 2, N depends only on Y.

$$N = \Gamma(B)(z/2)^{1-B} I_{B+1}(z)$$
$$z = \sqrt{16/3}(Y)$$
2)

The numerical value of dN/dy for several masses is shown in Fig.1. There is approximate mass independence at small y. The non-scaling behavior of the multiplicity appears as a growth in multiplicity as the mass increases which appears at large y or small x.

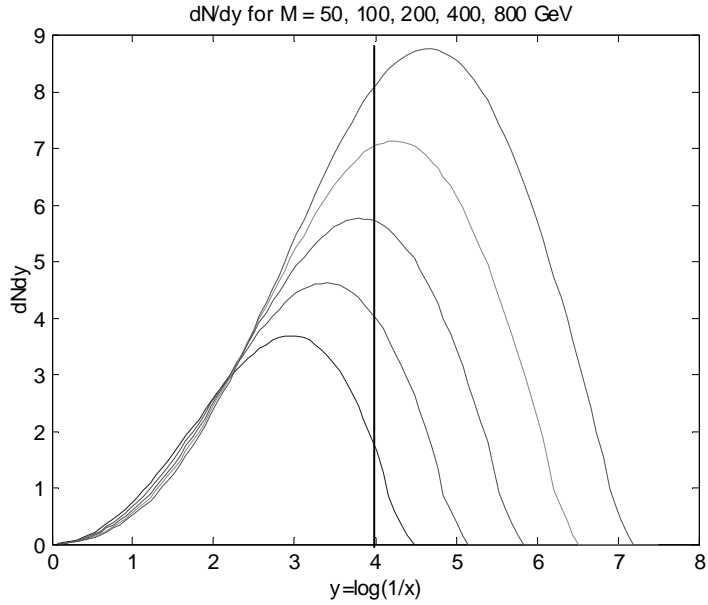

Figure 1: QCD prediction for the particle multiplicity density as a function of y for masses of 50, 100, 200, 400, and 800 GeV and for cone angle = 0.8. The vertical line indicates the cut, which is applied later.

The numerical values of the multiplicity for the masses used in Fig. 1 are shown in Fig. 2. At a mass of 120 GeV a mean multiplicity of ~ 12 hadrons is expected. Thus for the Z'(120) Monte Carlo events we expect a multiplicity ~ 12 and a y distribution which is fully contained for y < 5.

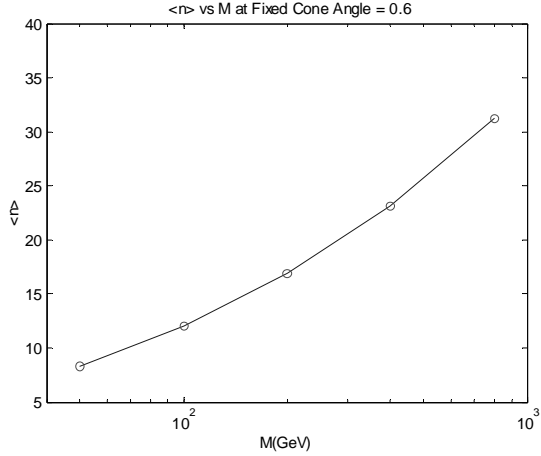

Figure 2: QCD prediction for the mean hadron multiplicity as a function of the dijet mass.

There is also the generic QCD prediction of angular ordering [7]. The "showering" partons are typically emitted at low momentum and small angle with respect to the jet. In a bit more detail, the partons emitted with a reasonably large x have small angles while those at very low x are expected to occur at larger angles. These QCD predictions will be used to guide the study made here in an effort to reduce the degradation in mass resolution due to pileup.

## Data Sets for Z'(120) at Low and High Luminosity

The CMS detector was fully simulated for the Drell-Yan production of Z' bosons with a mass of 120 GeV at design luminosity and at low (1/5) luminosity. A plot of the energy deposit in the calorimetry with the segmentation in rapidity and azimuth designed for the hadronic calorimetry is shown in Fig. 3 for single bunch crossings from the two data sets.

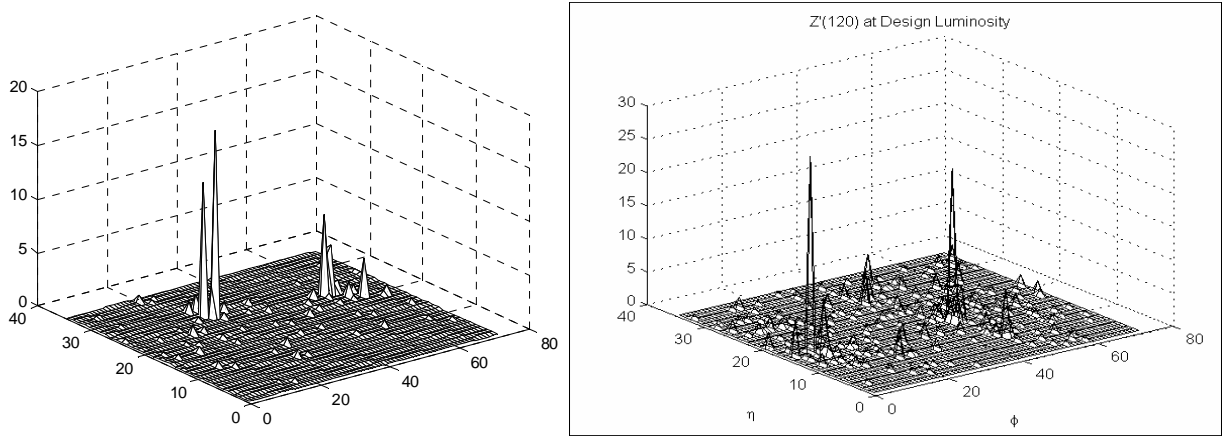

Figure 3: "LEGO" plots in pseudorapidity and azimuth, $(\eta, \phi), |\eta| < 1.4, 0 < \phi < 2\pi$ for a single bunch crossing at a) low luminosity and b) high luminosity. The granularity of the plot is that defined by the hadron calorimeter.

It is clear that at design luminosity the pileup effects must be carefully studied and mitigated if at all possible. Note, however, that even at design luminosity the "towers" of calorimetry are still somewhat sparsely populated.

At high luminosity the mean number of clusters within a bunch crossing as ~ 1000 for the calorimetry aperture of |y| < 5, or approximately 100 clusters (charged plus neutral) per unit of rapidity. If the mean number of interactions is 17 per crossing, that implies a density of about 5.9 clusters per unit of pseudorapidity per interaction. At low luminosity the mean number of calorimeter clusters is 240.

## Sufficient Cone Size

The data sets are used to find the jets in the "barrel" region, |y| < 1.4. The energy depositions in the calorimeters were searched first for "seeds" which exceed a given threshold in transverse momentum. A cone of radius Rc is then erected with the axis being the seed direction and all calorimeter clusters above a threshold Eth are added to the cone. The cone axes are recalculated by taking a mean of all clusters within the cone in both rapidity and azimuth weighted by transverse momentum, Et.

The resulting cones are then merged if their centers are within Rsep of one another. The merged cones are called jets. The jets have been taken to be massless.

At low luminosity a seed threshold of 2.5 GeV was found to be efficient. On average 9.8 seeds were found, 3.5 cones and 2.8 jets using a cone radius of Rc = 0.8 and Rsep = 1.6. There were on average 17 clusters in the highest Et cone. A background cone was set up at 90 degrees in azimuth to the leading cone. There were 6.6 clusters within it, leading to an estimate of 10.5 clusters associated with the jet itself. That number is roughly consistent with the MLLA prediction shown in Fig. 2.

At design luminosity the pileup is much worse and more care must be taken. Clearly, the pileup in number of included clusters within the cone increases as the square of the cone radius. Therefore, it is imperative to take as small a cone radius as possible. In Fig. 4 is shown the subtracted number of clusters in a cone as a function of the radius. The subtraction is again done using a cone of the same radius but at 90 degrees in azimuth away from the leading cone. Clearly, a radius of < 0.6 loses real particles. Therefore, Rc = 0.6 is the smallest cone that can be used without excluding real jet related particles.

The mean multiplicity within the leading cone at design luminosity is 28.7 clusters, while there are 19.1 clusters in the "background" cone at 90 degrees in azimuth to that cone for a cone radius of Rc= 0.6. The estimated mean number of jet related clusters is 9.6, roughly consistent with the result at low luminosity, which is encouraging.

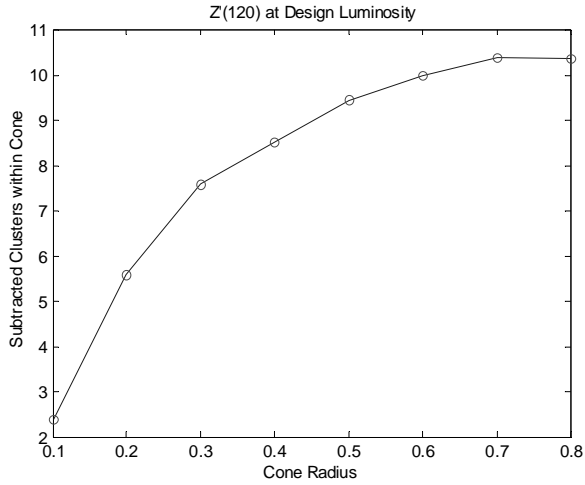

Figure 4: Mean number of clusters within a cone as a function of the radius of that cone. The mean has been corrected for pileup by the subtraction of the mean number of clusters within a cone of the same radius but at 90 degrees in azimuth to the leading jet direction.

## Defining Cuts in the (log(x), dR) Plane

Having studied the minimum cone radius and chosen Rc = 0.6, we know that this size cone contains a lot of pileup clusters which are not associated with the parton jet. The trivial ways to reduce the pileup are to use a smaller cone radius or to impose a threshold on the cluster transverse momentum. Smaller cone radii will lose jet particles (Fig. 4) and a threshold would cut away soft jet fragments (Fig. 1).

Therefore, cuts were devised in the (log(x),dR) plane. The variable log(x) is a natural QCD variable (Eq. 1).  In order to maximize the signal to noise, the low luminosity set was used for the studies that defined the cuts. The (log(x), dR) data at low luminosity were examined for both the leading cone and the "background" cone.  The low luminosity parameters of jet finding, Rc = 0.8, Rsep = 1.6 and Eth = 0.1 GeV were used.

There were very different populations of the plane for signal plus background and background. Cuts within the plane were developed to preferentially remove background. The cuts are designed to retain clusters with large transverse momentum which are cleanly distinct from the pileup background. That choice is driven by the knowledge that if the dijets are back to back then the mass of the dijet is determined largely by the scalar transverse momentum which exists within the jet cone.

$$\begin{aligned} M^2 &\sim 2(E_1 E_2 - \vec{P}_1 \cdot \vec{P}_2) \\ &\sim 4 E_1 E_2 \end{aligned} \qquad 3)$$

The "core" of the jet at small angles with respect to the jet axis and containing substantial values of x was very apparent in the signal and absent in the background. Clusters with ~ 2% or more of the jet energy ( $\log(x) > -4$ ) are retained – see Fig. 1. Clusters occurring at higher energy but with angles up to R ~ 0.56 are retained, as is necessary – see Fig. 4. A linear cut border extending from ($\log(x)$,R) of (-4,0.56) to (0,0.1) was adopted. These limits will not make major impacts on the dijet mass – see Eq. 3. With the cuts thus chosen there are 10.9 clusters on average in the leading cone, 5.2 in the background cone or 5.6 signal clusters retained out of the ~ 10 total (see Fig. 4).

The ($\log(x)$,dR) plane for the leading cone, containing both signal and pileup background, was examined after having had the background for pileup subtracted on average. The resulting plane population after this level of subtraction was also examined. The "core" of the jet remained after subtraction as did the softer fragments at small dR. There remained clusters with $\log(x)$ ~ -5 which are removed by the cuts. However, these clusters will not make a major effect on the dijet mass. Therefore, the cuts appear to be reasonable, on average.

The Z'(120) data for low luminosity does not have these cuts applied, because they clearly remove real jet clusters. Rather a simple and ample cone of radius Rc = 0.8 is used. The projections of the plane onto dR and $\log(x)$ for the Z'(120) data at design luminosity and a cone Rc = 0.6 (Fig. 4) are shown in Fig. 5. There is clearly a pileup background dN/dR ~ R for the background cone. For the leading cone the background and a signal peak at dR ~ 0.2 is evident. The background $\log(x)$ distribution has a single peak and is contained within $\log(x) < -2$. The leading cone displays a "core" for $\log(x) > -1$ and a peak at $\log(x)$ ~ -4.5 (see Fig. 1). Since the distributions are normalized, a background subtraction indicates that the leading cone also has clusters with small x values, $\log(x)$ ~ -4.

The Z'(120) data at low luminosity simply adopted the jet finding parameters, Rc = 0.8, Rsep = 1.6, and Eth = 0.1 GeV. For the design luminosity data set Rc = 0.6, Rsep = 1.2 and Eth = 0.1 GeV was used and the cuts which were developed were applied to the clusters within those cones.

## Dijet Mass Resolution

The resulting dijet mass distributions are shown in Fig. 6. Gaussian fits to the mass peaks resulted in the parameters of mean mass = 100.16 GeV (99.5 GeV) and sigma = 15.5 (15.0) GeV for the low and design luminosities. The error on the mass was ~ 1.4 GeV and on the sigma parameter was ~ 2.1 GeV, so that the resolution of the fractional mass is the same within errors at the two luminosities. However, there is clearly a high mass "tail" at design luminosity.

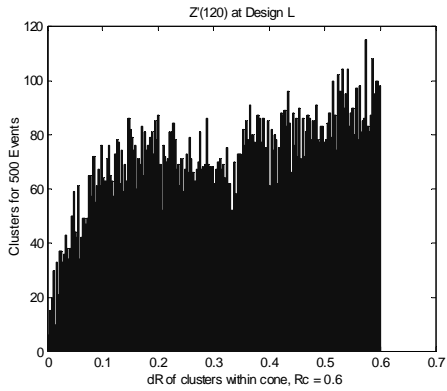

a)

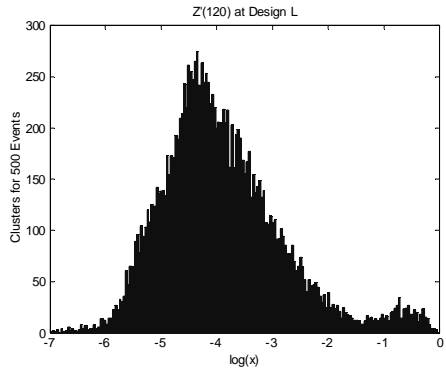

c)

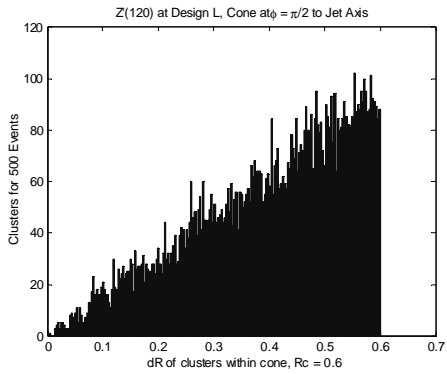

c)

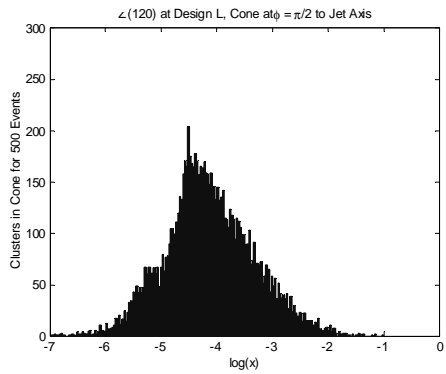

d)

Figure 5: Distribution of dR for Z'(120) data at design luminosity. a) leading cone b) "background" cone. Distribution of log(x) for c) leading cone and d) "background" cone.

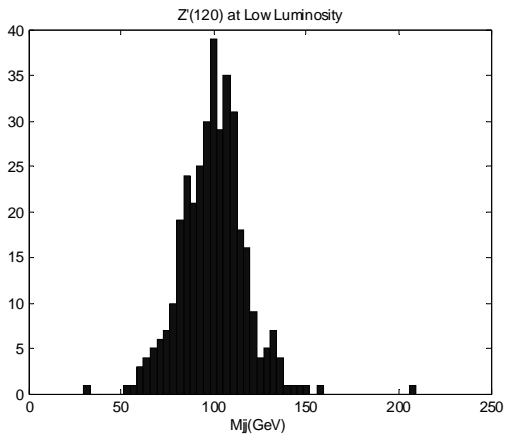

a)

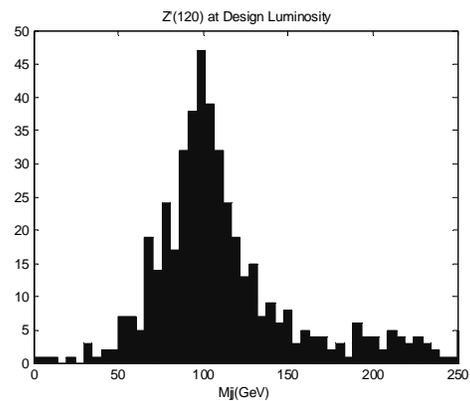

b)

Figure 6: Dijet mass distribution for Z'(120) data for a) low luminosity and b) design luminosity.

## Conclusions

It appears that a judicious set of cuts can be applied which evades a serious degradation of the mass resolution at design luminosity for LHC experiments. These cuts are motivated by QCD considerations of the characteristics of parton showers in the evolution of jets.